\documentclass[10pt]{article}

\usepackage{amsmath}
\usepackage{amssymb}

\usepackage{graphicx}

\usepackage{cite}

\usepackage{color}
\usepackage{setspace}
\usepackage[labelfont=bf,labelsep=period,justification=raggedright]{caption}

\makeatletter
\renewcommand{\@biblabel}[1]{\quad#1.}
\makeatother

\date{}

\begin{document}

\begin{flushleft}
{\Large
\textbf{Link Prediction in Complex Networks: A Mutual Information Perspective}
}
\\
Fei Tan,
Yongxiang Xia$^{\ast}$,
Boyao Zhu
\\
Department of Information Science and Electronic Engineering, Zhejiang University, Hangzhou, Zhejiang, China
\\
$\ast$ E-mail: xiayx@zju.edu.cn
\end{flushleft}

\section*{Abstract}

Topological properties of networks are widely applied to study the link-prediction problem recently. Common Neighbors, for example, is a natural yet efficient framework. Many variants of Common Neighbors have been thus proposed to further boost the discriminative resolution of candidate links. In this paper, we reexamine the role of network topology in predicting missing links from the perspective of information theory, and present a practical approach based on the mutual information of network structures. It not only can improve the prediction accuracy substantially, but also experiences reasonable computing complexity.

\section*{Introduction}

Link prediction attempts to estimate the likelihood of the existence of links between nodes based on the available network information, such as the observed links and nodes' attributes \cite{liben2007link,lu2011link}. On the one hand, the link-prediction problem is a long-standing practical scientific issue. It can find broad applications in both identifying missing and spurious links and predicting the candidate links that are expected to appear with the evolution of networks \cite{liben2007link,guimera2009missing,clauset2008hierarchical}. In biological networks (such as protein-protein interaction networks \cite{yu2008high} and metabolic networks \cite{jeong2000large} ), for example, the discovery of interactions is usually costly. Therefore, accurate prediction is more reasonable compared with blindly checking all latent interaction links \cite{guimera2009missing,clauset2008hierarchical}. In addition, the detection of inactive or anomalous connections in online social networks may improve the performance of link-based ranking algorithms \cite{zeng2012removing}. Furthermore, in online social networks \cite{kleinberg2013analysis}, very promising candidate links (non-connected node pairs) can be recommended to the relevant users as potential friendships. It can help them to find new friends and thus enhance their loyalties to the web sites. On the other hand, theoretically, link prediction can provide a useful methodology for the modeling of networks \cite{wang2012evaluating}. The evolving mechanisms of networks have been widely studied. Many evolving models have been proposed to capture the evolving process of real-world networks \cite{barabasi1999emergence,watts1998collective,albert2000topology,kumar2010structure}. However, it is very hard to quantify the degree to which the proposed evolving models govern real networks. Actually, each evolving model can be viewed as the corresponding predictor, we can thus apply evaluating metrics on prediction accuracy to measure the performance of different models.

Therefore, link prediction has attracted much attention from various scientific communities.
Within computer society, for example, scientists have employed Markov chains \cite{sarukkai2000link,zhu2002using} and machine learning techniques \cite{pavlov2007finding,benchettara2010supervised,soundarajan2012use,sa2010supervised,al2006link} to extract features of networks. These methods, however, depend on the attributes of nodes for particular networks such as social and textual features. Obviously, the attributes of nodes are generally hidden, and it is thus difficult for people to obtain them \cite{lu2011link}.

Over the last 15 years, network science has been developed as a novel framework for understanding structures of many real-world networked systems. Recently, a wealth of algorithms based on  structural information have been proposed \cite{lu2011link,newman2001clustering,leicht2006vertex,adamic2003friends,zhou2009predicting,liu2013hidden,liu2011link,cannistraci2013link,clauset2008hierarchical}.
Among various node-neighbor-based indices, Common Neighbors (CN) is undoubtedly the precursor with low computing complexity. It has also been revealed that CN achieves high prediction accuracy compared with other classical prediction indices \cite{zhou2009predicting}. CN, however, only emphasizes the number of common neighbors but ignores the difference in their contributions. In this case, several variants of CN to correct such a defect were put forwarded.
Consider, for example, Adamic-Adar \cite{adamic2003friends} and Resource Allocation \cite{zhou2009predicting}, in which low-degree common neighbors are advocated by assigning more weight to them. In addition, based on the Bayesian theory, a Local Na\"{\i}ve Bayes model \cite{liu2011link} was presented to differentiate the roles of neighboring nodes. Furthermore, node centrality (including degree, closeness and betweenness) was applied to make neighbors more distinguishable. Besides such CN-based indices, the evolving patterns and organizing principles of networks can also provide useful insights for coping with the link-prediction problem.  The well-known mechanism of preferential attachment \cite{barabasi1999emergence}, for instance, has been viewed as a prediction measure \cite{huang2005link,zhou2009predicting}.
For networks exhibiting hierarchical structure, Hierarchical Random Graph can be employed to predict missing links accordingly \cite {clauset2008hierarchical}. Recently, communities have been reinvented as groups of links rather than nodes \cite{ahn2010link}. Motivated by the shift in perspective of communities, Cannistraci et al. developed the local-community-paradigm to enhance the performance of classical prediction techniques \cite{cannistraci2013link}.

All the aforementioned methods aim to quantify the existence likelihood of candidate links. In information theory, the likelihood can be measured by the self-information. In this paper, we thus try to give a more theoretical analysis of the link-prediction problem from the perspective of information theory. Then a general prediction approach based on mutual information is presented accordingly. Our framework outperforms other prediction methods greatly.

\section*{Materials and Methods}

\subsection*{Data and Problem Description}

\begin{table*}[!ht]
\renewcommand{\arraystretch}{1.3}
\caption{The basic structural parameters of the giant components of example networks. $N$ and $M$ are the network size and the number of links, respectively. $e$ is the network efficiency \cite{latora2001efficient}, denoted as $e = \frac{2}{N(N-1)}\sum_{x,y \in V, x \neq y}d_{xy}^{-1}$, where $d_{xy}$ is the shortest distance between nodes $x$ and $y$. $C$ and $r$ are clustering coefficient \cite{watts1998collective} and assortative coefficient \cite{newman2002assortative}, respectively. $\langle k \rangle$ and $\langle d \rangle$ are the average degree and the average shortest distance. $H$ denotes the degree heterogeneity defined as $H = \frac {\langle k^2 \rangle}{{\langle k \rangle}^2}$.}
\label{Network}
\centering
\begin{tabular}{|c|c|c|c|c|c|c|c|c|}
\hline
Network $\setminus$ Index & N & M & e & C & r & H & $\langle k \rangle$ & $\langle d \rangle$ \\
\hline\hline
Email     & 1133 &  5451  & 0.2999  & 0.2540  &  0.0782  &  1.9421  &  9.6222  &  3.6028  \\
\hline
PB        & 1222 &  16714 &  0.3982 & 0.3600  &  -0.2213 &  2.9707  &  27.3552 &  2.7353 \\
\hline
Yeast     & 2375 &  11693 &  0.2181 & 0.3883  &  0.4539  &  3.4756  &  9.8467  &  5.0938 \\
\hline
SciMet    & 2678 &  10368 &  0.2569 & 0.2026  &  -0.0352 &  2.4265  &  7.7431  &  4.1781 \\
\hline
Kohonen   & 3704 &  12673 &  0.2957 & 0.3044  &  -0.1211 &  9.3170  &  6.8429  &  3.6693  \\
\hline
EPA       & 4253 &  8897  &  0.2356 & 0.1360  &  -0.3041 &  6.7668  &  4.1839  &  4.4993      \\
\hline
Grid      & 4941 &  6594  &  0.0629 & 0.1065  &  0.0035  &  1.4504  &  2.6691  &  18.9853 \\
\hline
INT       & 5022 &  6258  &  0.1667 & 0.0329  &  -0.1384 &  5.5031  &  2.4922  &  6.4475     \\
\hline
Wikivote  & 7066 &  100736&  0.3268 & 0.2090  &  -0.0833 &  5.0992  &  28.5129 &  3.2471       \\
\hline
Lederberg & 8212 &  41430 &  0.2560 & 0.3634  &  -0.1001 &  6.1339  &  10.0901 &   4.4071       \\
\hline
\end{tabular}
\end{table*}

In this paper, in order to reduce the randomness caused by the network size, we choose ten example data sets from various areas with the size of its giant component being greater than 1000. They are listed as follows.
i) Email \cite{dong2011link}: A network of Alex Arenas's email. ii) PB \cite{adamic2005political}: A network of the US political blogs. iii) Yeast \cite{von2002comparative}: A protein-protein interaction network. iv) SciMet \cite{Pajek}: A network of articles from or citing Scientometrics. v) Kohonen \cite{Pajek}: A network of articles with topic ¡°self-organizing maps¡± or references to ¡°Kohonen T¡±. vi) EPA \cite{EPA}: A network of web pages linking to the website www.epa.gov. vii) Grid \cite{watts1998collective}: An electrical power grid of the western US. viii) INT \cite{spring2002measuring}: The router-level topology of the Internet. ix) Wikivote \cite{leskovec2010predicting,leskovec2010signed}: The network contains all the Wikipedia voting data from the inception of Wikipedia till January 2008. x) Lederberg \cite{Lederberg}: A network of articles by and citing J. Lederberg, during the year 1945 to 2002. Here we only focus on the giant component of networks. Their basic topological parameters are summarized in Table \ref{Network}.

In this paper, only an undirected simple network $G(V,E)$ is studied, where $V$ and $E$ are the sets of nodes and of links, respectively. That is to say, the direction of links, self-connections and multiple links are ignored here. The framework of prediction indices can be described as follows \cite{lu2011link}. Given a disconnected node pair $(x,y)$, where $x,y \in V$, we should try to predict the likelihood of connectivity between them. For each non-existent link $(x,y) \in U - E$, where $U$ represents the universal set, a score $s_{xy}$ will be given to measure its existence likelihood according to a specific predictor. The higher the score is, the more possible the node pair has a candidate link. To figure out the latent links, all disconnected ones are first sorted in the descending order. The top-ranked node pairs are believed most likely to have links.

To validate the prediction performance of the algorithms, the observable links of the network are divided into two separate sets, i.e., the training set $E^T$ and the probe set $E^P$. Obviously, $E^T$ is the available topological information, and $E^P$ is for the test and thus cannot be used for prediction. Therefore, $E^T \cup E^P = E$ and $E^T \cap E^P = \emptyset$. In our model, the training set $E^T$ and probe set $E^P$ are assumed to contain $90\%$ and $10\%$ of links, respectively (see the review article \cite{lu2011link} and references therein).

As in many previous papers, two widely used metrics are adopted to evaluate the performance of prediction algorithms \cite{lu2011link}. They are AUC (area under the receiver operating characteristic curve) \cite{hanley1982ROC} and precision \cite{herlocker2004evaluating}. AUC is denoted as follows:
\begin{equation}
AUC = \frac{n{'} + 0.5n^{''}}{n},
\end{equation}
where among $n$ times of independent comparisons, $n{'}$ and $n{''}$ represent the time that a randomly chosen missing link has a higher score and the time that they share the same score compared with a randomly chosen nonexistent link, respectively. Clearly, AUC should be around $0.5$ if all scores follow an independent and identical distribution. Therefore, as a macroscopic accuracy measure, the extent to which AUC exceeds 0.5 indicates the performance of a specific method compared with pure chance. Another popular measure is precision, which focuses on top-ranked latent links. It is defined as $L_r/L$, where among top-$L$ candidate links, $L_r$ is the number of accurate predicted links in the probe set.

\subsection*{Previous Prediction Methods}

We here introduce six typical methods based on common neighbors. They are Common Neighbors (CN), Resource Allocation (RA) \cite{zhou2009predicting}, the Local Na\"{\i}ve Bayes (LNB) forms of CN  \cite{liu2011link} and RA \cite{liu2011link}, CAR \cite{cannistraci2013link} and CRA \cite{cannistraci2013link}, respectively. We denote the set of node $x$'s neighboring nodes by $\Gamma(x)$. For node pair $(x,y)$, the set of their common neighbors is denoted as $O_{xy}=\Gamma(x) \cap \Gamma(y)$.

\begin{itemize}
  \item  CN. This method is the natural framework in which the more nodes $x$ and $y$ share common neighbors, the more likely they are connected. The score can be quantified by the number of their common neighbors, namely
          \begin{equation}
          s_{xy}^{CN} = |\Gamma(x) \cap \Gamma(y)| = |O_{xy}|.
          \end{equation}

    \item  RA. In this method, the weight of the neighboring node is negatively proportional to its degree. The score is thus denoted as
    \begin{equation}
     s_{xy}^{RA} = \sum_{z \in O_{xy}}\frac{1}{|\Gamma(z)|}.
    \end{equation}

   \item LNB-CN. Based on the na\"{\i}ve Bayes classifier, this method combines CN and the clustering coefficient together. The score is defined as
          \begin{equation}
          s_{xy}^{LNB-CN} = |O_{xy}|\log{\eta} + \sum_{z\in{O_{xy}}}\log{R_z}.
          \end{equation}

          In this formula, $\eta$ is denoted as
          \begin{equation}
          \eta = \frac{|V|(|V|-1)}{2|E^T|} - 1.
          \end{equation}

          In addition, $R_z$ is defined as
          \begin{equation}
          R_z= \frac{N_{\triangle z} + 1}{N_{\wedge z} + 1},
          \end{equation}
          where $N_{\triangle z}$ and $N_{\wedge z}$ are the numbers of connected and of disconnected node pairs with node $z$ being a common neighbor, respectively.

   \item LNB-RA. Similarly to LNB-CN, this method takes RA and the clustering coefficient into account. The score is thus denoted as
          \begin{equation}
          s_{xy}^{LNB-RA} = \sum_{z\in{O_{xy}}}\frac{1}{|\Gamma(z)|}(\log{\eta} + \log{R_z}).
          \end{equation}

  \item  CAR. This method boosts the discriminative resolution between latent links characterized by the same number of common neighbors through further emphasizing the link community among such common neighbors. Thus, it is described as \begin{equation}
      s_{xy}^{CAR} = |O_{xy}| \cdot \sum_{z \in O_{xy}}\frac{|\gamma(z)|}{2},
      \end{equation}
      where $\gamma(z)$ refers to the subset of neighbors of node $z$ that are also common neighbors of nodes $x$ and $y$.

   \item CRA. This method is a variation of CAR when RA is considered. It can be thus denoted as
          \begin{equation}
          s_{xy}^{CRA} = \sum_{z \in O_{xy}}\frac{|\gamma(z)|}{|\Gamma(z)|}.
          \end{equation}

\end{itemize}

\subsection*{A Mutual Information Approach to Link Prediction}

We here introduce the definitions of the self-information and of the mutual information, respectively.

\textbf{Definition 1} Considering a random variable $X$ associated with outcome $x_k$ with probability $p(x_k)$, its {\it self-information} $I(x_k)$ can be denoted as \cite{shannon2001mathematical}
\begin{equation}
I(x_k)= \log \frac{1}{p(x_k)}= -\log p(x_k),
\end{equation}
where the base of the logarithm is specified as 2, thus the unit of self-information is bit. This is applicable for the following if not otherwise specified. The self-information indicates the uncertainty of the outcome $x_k$. Obviously, the higher the self-information is, the less likely the outcome $x_k$ occurs.

\textbf{Definition 2} Consider two random variables $X$ and $Y$ with a joint probability mass function $p(x,y)$ and marginal probability mass functions $p(x)$ and $p(y)$. The { \it mutual information} $I(X;Y)$ can be denoted as follows \cite{cover2012elements}:
\begin{equation}
\begin{split}
I(X;Y) &= \sum_{x \in X}\sum_{y \in Y}p(x,y)\log \frac{p(x,y)}{p(x)p(y)}\\
&= \sum_{x,y}p(x,y)\log \frac{p(x,y)}{p(x)p(y)}\\
&= \sum_{x,y}p(x,y)\log \frac{p(x|y)}{p(x)}.
\label{Def1}
\end{split}
\end{equation}
Hence, the mutual information $I(x_k;y_j)=I(X=x_k;Y=y_j)$ can be obtained as
\begin{equation}
\begin{split}
I(x_k;y_j) & = \log \frac{p(x_k|y_j)}{p(x_k)}\\
&= -\log p(x_k)-(-\log p(x_k|y_j))\\
&= I(x_k) - I(x_k|y_j).
\label{Def2}
\end{split}
\end{equation}
The mutual information is the reduction in uncertainty due to another variable. Thus, it is a measure of the dependence between two variables. It is equal to zero if and only if two variables are independent.

Now given node $z$, assume there are some links among its neighboring nodes. According to eq. (12), the mutual information between the event that node $z$ is the common neighbor of two randomly chosen nodes and the event that they are connected can be denoted as follows:
\begin{equation}
I(L^1;z) = I(L^1) - I(L^1|z),
\label{1}
\end{equation}
where $I(L^1)$ is the self-information of that a randomly chosen node pair has one link. $I(L^1|z)$ is the conditional self-information of that a node pair is connected when their common neighbors include node $z$.
Particularly, $p(L^1|z)$ is the clustering coefficient of node $z$, defined as
\begin{equation}
p(L^1|z) = \frac{N_{\triangle z}}{N_{\triangle z} + N_{\wedge z}},
\end{equation}
where $N_{\triangle z}$ and $N_{\wedge z}$ are as same as those in eq. (6).

As we here mainly focus on the neighbors of node $z$, $I(L^1)$ is averaged over the corresponding self-information of all neighboring node pairs. It can be thus denoted as
\begin{equation}
I(L^1) = \frac{1}{|\Gamma(z)|(|\Gamma(z)| - 1)}\sum_{m \neq n \atop m,n \in \Gamma(z)}I(L_{mn}^1),
\end{equation}
where $I(L_{mn}^1)$ is the self-information of that node pair $(m,n)$ has one link.

We assume that no degree-degree correlation is considered. When nodes' degree is only known, the probability that node pair $(m,n)$ is disconnected can be derived as
\begin{equation}
\begin{split}
p(L_{mn}^0) &= \prod_{i=1}^{k_n}\frac{(M-k_m)-i+1}{M-i+1} \\
& = \frac{C_{M-k_m}^{k_n}}{C_M^{k_n}},
\end{split}
\end{equation}
where $k_m$ and $k_n$ are the degrees of nodes $m$ and $n$, respectively. $M$ is the total number of links in the training set. Obviously, this formula is symmetric, namely
\begin{equation}
p(L_{nm}^0) = \frac{C_{M-k_n}^{k_m}}{C_M^{k_m}} = \frac{C_{M-k_m}^{k_n}}{C_M^{k_n}} = p(L_{mn}^0)   .
\end{equation}

Thus,
\begin{equation}
\begin{split}
p(L_{nm}^1) = p(L_{mn}^1)
= 1 - \frac{C_{M-k_m}^{k_n}}{C_M^{k_n}}.
\end{split}
\end{equation}

Collecting these results, we can obtain \\
\\

\begin{equation}
\begin{split}
{I(L^1;z)} &= \frac{1}{|\Gamma(z)|(|\Gamma(z)| - 1)}\sum_{m \neq n \atop m,n \in \Gamma(z)}-\log{p(L_{mn}^1)} - (-\log{p(L^1|z)})
\\
&= \frac{1}{|\Gamma(z)|(|\Gamma(z)| - 1)}\sum_{m \neq n \atop m,n \in \Gamma(z)}\log{\frac{C_M^{k_n}}{C_M^{k_n} -C_{M-k_m}^{k_n}}}
+\log{\frac{N_{\triangle z}}{N_{\triangle z} + N_{\wedge z}}},
\end{split}
\end{equation}
where it is stipulated that $I(L^1;z)=0$ if $N_{\triangle z} = 0$.

Now given a disconnected node pair $(x,y)$, if the set of their common neighbors $O_{xy}$ is available, the self-information of the existence of a link between them can be derived as
\begin{equation}
I(L_{xy}^1|O_{xy}) = I(L_{xy}^1) - I(L_{xy}^1;O_{xy}),
\end{equation}
where $I(L_{xy}^1)$ can be calculated according to eq. (18). If the elements of $O_{xy}$ are assumed to be independent of each other, then
\begin{equation}
I(L_{xy}^1;O_{xy}) = \sum_{z \in O_{xy}}I(L_{xy}^1;z),
\end{equation}
where node $z$ is the common neighbor of nodes $x$ and $y$. $I(L_{xy}^1;z)$ can thus be estimated by $I(L^1;z)$. We substitute eq. (21) into eq. (20) and obtain
\begin{equation}
I(L_{xy}^1|O_{xy}) = I(L_{xy}^1) - \sum_{z \in O_{xy}}I(L^1;z).
\end{equation}
As we utilize the mutual information of common neighbors to estimate the connection likelihood, this framework is called{ \it MI} for short. According to the property of the self-information, the smaller $I(L_{xy}^1|O_{xy})$ is, the higher the likelihood of existence of links is. Thus, we denote the score as
\begin{equation}
\begin{split}
s_{xy}^{MI} & = -I(L_{xy}^1|O_{xy}) \\
& = \sum_{z \in O_{xy}}I(L^1;z) - I(L_{xy}^1).
\end{split}
\end{equation}

\begin{figure}[!ht]
\centering
\includegraphics[height=0.5\columnwidth,width=0.5\columnwidth]{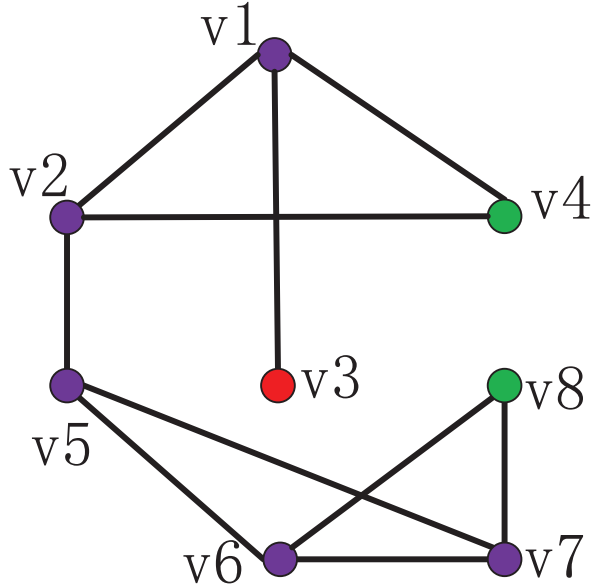}
\caption{(Color online) An illustration about the calculation of MI model.}
\label{MI}
\end{figure}

To facilitate the understanding of MI, we illustrate it with an example as shown in fig. \ref{MI}. First, consider node $v1$, for example, which is the common neighbor of nodes $v2$, $v3$ and $v4$. Using eq. (14), we can have $I(L^1|v1) = \log{3}=1.585$. Based on eq. (18), we obtain $I(L_{v2v3}^1) = \log{\frac{10}{3}}=1.737$, $I(L_{v2v4}^1) = \log{\frac{15}{8}}=0.9069$ and $I(L_{v3v4}^1) = \log5 = 2.3219$. Hence, we have $I(L^1;v1)=0.0703$. Now we compare node pairs $(v2,v3)$ and $(v3,v4)$ with the common neighbor node $v1$. Then $I(L^1_{v2v3}|v1)=1.6667$, $I(L^1_{v3v4}|v1)=2.2516$, which can be calculated based on eq. (22). That is to say, node pair $(v2,v3)$ is more likely to be connected than node pair $(v3,v4)$. The above-mentioned six prediction methods, however, cannot distinguish these two node pairs. In this sense, MI has higher discriminative resolution than them. Second, MI can distinguish node pairs even if they all have no common neighbors. For instance, $I(L_{v3v5}^1)=\log{\frac{10}{3}}=1.7370$ and $I(L_{v3v8}^1) = \log5 = 2.3219$. That is to say, node pair $(v3,v5)$ is more likely to be connected than node pair $(v3, v8)$. This is undoubtedly beyond the distinguishing ability of previous methods. Thirdly, the mutual information of node $v6$ can be calculated as $I(L^1;v6)= I(L^1;v7) = 0.1854$. Thus $I(L^1_{v5v8}|v6v7)=0.5361$. We note that $I(L^1_{v5v8}|v6v7)<I(L^1_{v2v3}|v1)$, namely, node pair $(v5,v8)$ with two common neighbors has higher connection likelihood compared to node pair $(v2,v3)$ with only one common neighbor. This is in agreement with our intuition very well. Lastly, different nodes may provide different mutual information to reduce the uncertainty of connections. The extent to which node $v6$ ($I(L^1;v6)=0.1854$) contributes to the reduction of link uncertainty, for example, is greater than that of node $v1$ ($I(L^1;v1)=0.0703$).

\section*{Results and Discussion}

\begin{table*}[!t]
\renewcommand{\arraystretch}{1.3}
\caption{Comparison of the prediction accuracy measured by AUC on ten real-world networks. Each value is averaged over 100 independent runs with random divisions of training set $(90\%)$ and probe set $(10\%)$. The bold font represents that MI is better than the corresponding prediction index.}
\label{AUC}
\centering
\begin{tabular}{|c|c|c|c|c|c|c|c|}
\hline
Network $\setminus$ Index  & CN & RA & LNB-CN & LNB-RA & CAR & CRA & MI  \\
\hline\hline
Email     &\textbf{0.8574}	&\textbf{0.8592}	&\textbf{0.8588}	&\textbf{0.8592}	&\textbf{0.7039}	&\textbf{0.7042}	 &0.8917    \\
\hline
PB        &\textbf{0.9233}	&\textbf{0.9286}	&\textbf{0.9263}	&\textbf{0.9284}	&\textbf{0.896}	&\textbf{0.8976}	 &0.9322   \\
\hline
Yeast     &\textbf{0.9157}	&\textbf{0.9167}	&\textbf{0.9162}	&\textbf{0.9165}	&\textbf{0.8473}	&\textbf{0.8476}	 &0.9368   \\
\hline
SciMet    &\textbf{0.7997 }	&\textbf{0.8008 }	&\textbf{0.8013 }	&\textbf{0.8013 }	&\textbf{0.6131 }	&\textbf{0.6129 }	 &0.871   \\
\hline
Kohonen   &\textbf{0.8272}	&\textbf{0.8344}	&\textbf{0.8349}	&\textbf{0.835}	&\textbf{0.6489}	&\textbf{0.6493}	 &0.9111    \\
\hline
EPA       &\textbf{0.6118}	&\textbf{0.6131}	&\textbf{0.6139}	&\textbf{0.6138}	&\textbf{0.508}	&\textbf{0.5079}	 &0.9249        \\
\hline
Grid      &0.6257	&0.6255	&0.6258	&0.6256	&\textbf{0.517}	&\textbf{0.5171}	&0.6076   \\
\hline
INT       &\textbf{0.6523}	&\textbf{0.6526}	&\textbf{0.6523}	&\textbf{0.6525}	&\textbf{0.5277}	&\textbf{0.5281}	 &0.9559       \\
\hline
Wikivote  &\textbf{0.9389}	&\textbf{0.94}	&\textbf{0.9401}	&\textbf{0.9398}	&\textbf{0.8899}	&\textbf{0.8909}	 &0.9663        \\
\hline
Lederberg & \textbf{0.9024}	&\textbf{0.9058}	&\textbf{0.9061}	&\textbf{0.9061}	&\textbf{0.7417}	&\textbf{0.7414}	 &0.9449        \\
\hline
\end{tabular}
\end{table*}

\begin{table*}[!t]
\renewcommand{\arraystretch}{1.3}
\caption{Comparison of the prediction accuracy measured by precision (top-100) on ten real-world networks. Each value is averaged over 100 independent runs with random divisions of training set $(90\%)$ and probe set $(10\%)$. The bold font represents that MI is better than the corresponding prediction index.}
\label{Precision}
\centering
\begin{tabular}{|c|c|c|c|c|c|c|c|}
\hline

Network $\setminus$ Index  & CN & RA & LNB-CN & LNB-RA & CAR & CRA & MI  \\
\hline\hline
Email   &\textbf{0.3002}	&\textbf{0.2614}	&\textbf{0.3236}	&\textbf{0.2356}	&\textbf{0.3171}	&0.3442	&0.3293     \\
\hline
PB     &\textbf{0.4237}	&\textbf{0.2536}	&\textbf{0.414 }	&\textbf{0.2588}	&0.4795	&0.4876	&0.4765    \\
\hline
Yeast  &\textbf{0.6784}	&\textbf{0.4989}	&\textbf{0.6826}	&\textbf{0.5762}	&\textbf{0.6669}	&\textbf{0.7664}	 &0.8264      \\
\hline
SciMet  &\textbf{0.1411}	&\textbf{0.1265}	&\textbf{0.1511}	&\textbf{0.126 }	&0.1707	&0.1791	&0.166   \\
\hline
Kohonen   &\textbf{0.1577}	&\textbf{0.1435}	&\textbf{0.1698}	&\textbf{0.1462}	&\textbf{0.2097}	&0.2345	&0.224   \\
\hline
EPA   &\textbf{0.0156}	&\textbf{0.0375}	&\textbf{0.0277}	&\textbf{0.0398}	&\textbf{0.0271}	&\textbf{0.0546}	 &0.0578            \\
\hline
Grid    &\textbf{0.1161}	&\textbf{0.0866}	&\textbf{0.1604}	&\textbf{0.0968}	&\textbf{0.1255}	&0.1846	&0.1749  \\ \hline
INT  &\textbf{0.1021}	&\textbf{0.0869}	&\textbf{0.1221}	&\textbf{0.0636}	&\textbf{0.0829}	&\textbf{0.1247}	 &0.217           \\
\hline
Wikivote  &\textbf{0.189 }	&\textbf{0.1565}	&\textbf{0.1875}	&\textbf{0.1597}	&0.2639	&0.2849	&0.1933      \\
\hline
Lederberg    &\textbf{0.2402}	&\textbf{0.2958}	&\textbf{0.2606}	&\textbf{0.3075}	&\textbf{0.2699}	&0.3422	 &0.3312   \\
\hline
\end{tabular}
\end{table*}

In this section, we compare our MI approach with previous six representative prediction indices. Tables \ref{AUC} and \ref{Precision} show the prediction accuracy measured by AUC and precision, respectively. The overall prediction performance of MI outperforms them greatly.

Table \ref{AUC} demonstrates that for AUC, MI model gives much higher prediction accuracy than all 6 other indices for real-world networks except network Grid. Especially for networks EPA and INT, AUC of six indices is all around 0.6. MI model can experience AUC of more than 0.9. Such great difference may arise from that previous methods can't distinguish those node pairs without common neighbors. Unfortunately, the lack of common neighbors between two nodes often appear in real-world networks.
For example, more than 99\% of node pairs in network INT have no common neighbors. But MI approach is able to discriminate them greatly. Another finding is that CAR-based indices (CAR and CRA) achieve the worst prediction performance for ten networks. Actually, for node pairs with few common neighbors, the distinguishing ability of CAR-based indices degenerates remarkably due to their emphasis on the links among common neighbors. For example, all node pairs with less than two common neighbors share the same connection likelihood because they all have no links among common neighbors.

Table \ref{Precision} shows the comparisons of precision for ten real-world networks. We can see that MI is much better than CN, RA, LBN-CN, and LNB-RA for all networks. CAR-based indices, however, achieve higher precision than MI for some networks. The efficiency of CAR-based indices in predicting top-ranked candidate links is very high for networks with notable link communities. Consider, for example, network Wikivote with high average degree, in which CAR-based indices overwhelmingly win MI and other methods. Obviously, the extent to which CAR-based indices excel MI is positively related to link communities. The computing complexity of CAR-based indices, however, depends on the density of networks greatly.

It is thus necessary to compare the computing complexity of CAR-based indices and our MI model. Here the average degree is denoted as $\langle k \rangle$. According to eq. (8), the time complexity of computing $\gamma(z)$ and $O_{xy}$ is $O(\langle k \rangle^{4})$ and $O(\langle k \rangle ^ {2})$, respectively. The total computing complexity of CAR is thus $O(N^{2}\cdot\langle k \rangle ^{6})$. Similarly to CAR, the computing complexity of CRA is also $O(N^{2}\cdot\langle k \rangle ^{6})$ because $\Gamma(z)$ has the computing complexity of $O(1)$ based on eq. (9). For MI, the computing complexity of $I(L_{mn} ^{1})$  and averaging all neighboring node pairs of node $z$ is both $O(\langle k \rangle ^ {2})$. Thus, $I(L^1;z)$ has the computing complexity of $O(\langle k \rangle ^ {4})$. The computing complexity of MI model can be derived as $O(N^{2}\cdot\langle k \rangle ^ {4})$ accordingly. Taking precision and the computing complexity of CAR-based indices together, we note that they outperform MI in some networks but with the computing complexity as $\langle k \rangle ^ {2}$ times as that of MI. It is intolerable especially for networks with the high average degree. The average degree of network Wikivote, for instance, is $\langle k \rangle > 28$ (see Table \ref{Network}). This means that the order of magnitude of CAR-based indices' computing complexity is about three more than that of MI model.

Altogether, MI has a good tradeoff among AUC, precision and the computing complexity.

\section*{Conclusions}

In this paper, we develop a novel framework to uncover missing edges in networks via the mutual information of network topology. Note that our approach differs crucially from previous prediction methods in that it is derived strictly from information theory. We compare our model with six typical prediction indices on ten networks from disparate fields. The simulation results show that MI model overwhelms them. Furthermore, we compare the computing complexity of MI model with that of CAR-based indices and find that our approach is less time-consuming.

\section*{Acknowledgments}
This work was supported by the National Natural Science Foundation of China under Grant No. 61174153.

\bibliography{Link_Prediction}

\providecommand{\noopsort}[1]{}\providecommand{\singleletter}[1]{#1}%
\begin{thebibliography}{10}
\providecommand{\url}[1]{\texttt{#1}}
\providecommand{\urlprefix}{URL }
\expandafter\ifx\csname urlstyle\endcsname\relax
  \providecommand{\doi}[1]{doi:\discretionary{}{}{}#1}\else
  \providecommand{\doi}{doi:\discretionary{}{}{}\begingroup
  \urlstyle{rm}\Url}\fi
\providecommand{\bibAnnoteFile}[1]{%
  \IfFileExists{#1}{\begin{quotation}\noindent\textsc{Key:} #1\\
  \textsc{Annotation:}\ \input{#1}\end{quotation}}{}}
\providecommand{\bibAnnote}[2]{%
  \begin{quotation}\noindent\textsc{Key:} #1\\
  \textsc{Annotation:}\ #2\end{quotation}}
\providecommand{\eprint}[2][]{\url{#2}}

\bibitem{liben2007link}
Liben-Nowell D, Kleinberg J (2007) The link-prediction problem for social
  networks.
\newblock J Am Soc Inf Sci Technol 58: 1019--1031.
\bibAnnoteFile{liben2007link}

\bibitem{lu2011link}
L{\"u} L, Zhou T (2011) Link prediction in complex networks: A survey.
\newblock Physica A 390: 1150--1170.
\bibAnnoteFile{lu2011link}

\bibitem{guimera2009missing}
Guimer{\`a} R, Sales-Pardo M (2009) Missing and spurious interactions and the
  reconstruction of complex networks.
\newblock Proc Natl Acad Sci USA 106: 22073--22078.
\bibAnnoteFile{guimera2009missing}

\bibitem{clauset2008hierarchical}
Clauset A, Moore C, Newman ME (2008) Hierarchical structure and the prediction
  of missing links in networks.
\newblock Nature 453: 98--101.
\bibAnnoteFile{clauset2008hierarchical}

\bibitem{yu2008high}
Yu H, Braun P, Y{\i}ld{\i}r{\i}m MA, Lemmens I, Venkatesan K, et~al. (2008)
  High-quality binary protein interaction map of the yeast interactome network.
\newblock Science 322: 104--110.
\bibAnnoteFile{yu2008high}

\bibitem{jeong2000large}
Jeong H, Tombor B, Albert R, Oltvai ZN, Barab{\'a}si AL (2000) The large-scale
  organization of metabolic networks.
\newblock Nature 407: 651--654.
\bibAnnoteFile{jeong2000large}

\bibitem{zeng2012removing}
Zeng A, Cimini G (2012) Removing spurious interactions in complex networks.
\newblock Phys Rev E 85: 036101.
\bibAnnoteFile{zeng2012removing}

\bibitem{kleinberg2013analysis}
Kleinberg J (2013) Analysis of large-scale social and information networks.
\newblock Phil Trans R Soc A 371: 20120378.
\bibAnnoteFile{kleinberg2013analysis}

\bibitem{wang2012evaluating}
Wang WQ, Zhang QM, Zhou T (2012) Evaluating network models: A likelihood
  analysis.
\newblock EPL 98: 28004.
\bibAnnoteFile{wang2012evaluating}

\bibitem{barabasi1999emergence}
Barab{\'a}si AL, Albert R (1999) Emergence of scaling in random networks.
\newblock Science 286: 509--512.
\bibAnnoteFile{barabasi1999emergence}

\bibitem{watts1998collective}
Watts DJ, Strogatz SH (1998) Collective dynamics of small-world networks.
\newblock Nature 393: 440--442.
\bibAnnoteFile{watts1998collective}

\bibitem{albert2000topology}
Albert R, Barab{\'a}si AL (2000) Topology of evolving networks: local events
  and universality.
\newblock Phys Rev Lett 85: 5234.
\bibAnnoteFile{albert2000topology}

\bibitem{kumar2010structure}
Kumar R, Novak J, Tomkins A (2010) Structure and evolution of online social
  networks.
\newblock In: Link Mining: Models, Algorithms, and Applications, Springer. pp.
  337--357.
\bibAnnoteFile{kumar2010structure}

\bibitem{sarukkai2000link}
Sarukkai RR (2000) Link prediction and path analysis using markov chains.
\newblock Comput Netw 33: 377--386.
\bibAnnoteFile{sarukkai2000link}

\bibitem{zhu2002using}
Zhu J, Hong J, Hughes JG (2002) Using markov models for web site link
  prediction.
\newblock In: Proceedings of the Thirteenth ACM Conference on Hypertext and
  Hypermedia. ACM, pp. 169--170.
\bibAnnoteFile{zhu2002using}

\bibitem{pavlov2007finding}
Pavlov M, Ichise R (2007) Finding experts by link prediction in co-authorship
  networks.
\newblock FEWS 290: 42--55.
\bibAnnoteFile{pavlov2007finding}

\bibitem{benchettara2010supervised}
Benchettara N, Kanawati R, Rouveirol C (2010) Supervised machine learning
  applied to link prediction in bipartite social networks.
\newblock In: Proceedings of the International Conference on Advances in Social
  Network Analysis and Mining. IEEE, pp. 326--330.
\bibAnnoteFile{benchettara2010supervised}

\bibitem{soundarajan2012use}
Soundarajan S, Hopcroft JE (2012) Use of supervised learning to predict
  directionality of links in a network.
\newblock In: Advanced Data Mining and Applications, Springer. pp. 395--406.
\bibAnnoteFile{soundarajan2012use}

\bibitem{sa2010supervised}
S{\'a} HR, Prud{\^e}ncio RB (2010) Supervised learning for link prediction in
  weighted networks.
\newblock In: III International Workshop on Web and Text Intelligence.
\bibAnnoteFile{sa2010supervised}

\bibitem{al2006link}
Al~Hasan M, Chaoji V, Salem S, Zaki M (2006) Link prediction using supervised
  learning.
\newblock In: SDM¡¯06: Workshop on Link Analysis, Counter-terrorism and
  Security.
\bibAnnoteFile{al2006link}

\bibitem{newman2001clustering}
Newman ME (2001) Clustering and preferential attachment in growing networks.
\newblock Phys Rev E 64: 025102.
\bibAnnoteFile{newman2001clustering}

\bibitem{leicht2006vertex}
Leicht E, Holme P, Newman ME (2006) Vertex similarity in networks.
\newblock Phys Rev E 73: 026120.
\bibAnnoteFile{leicht2006vertex}

\bibitem{adamic2003friends}
Adamic LA, Adar E (2003) Friends and neighbors on the web.
\newblock Social Networks 25: 211--230.
\bibAnnoteFile{adamic2003friends}

\bibitem{zhou2009predicting}
Zhou T, L{\"u} L, Zhang YC (2009) Predicting missing links via local
  information.
\newblock Eur Phys J B 71: 623--630.
\bibAnnoteFile{zhou2009predicting}

\bibitem{liu2013hidden}
Liu H, Hu Z, Haddadi H, Tian H (2013) Hidden link prediction based on node
  centrality and weak ties.
\newblock EPL 101: 18004.
\bibAnnoteFile{liu2013hidden}

\bibitem{liu2011link}
Liu Z, Zhang QM, L{\"u} L, Zhou T (2011) Link prediction in complex networks: A
  local na{\"\i}ve bayes model.
\newblock EPL 96: 48007.
\bibAnnoteFile{liu2011link}

\bibitem{cannistraci2013link}
Cannistraci CV, Alanis-Lobato G, Ravasi T (2013) From link-prediction in brain
  connectomes and protein interactomes to the local-community-paradigm in
  complex networks.
\newblock Sci Rep 3: 1613.
\bibAnnoteFile{cannistraci2013link}

\bibitem{huang2005link}
Huang Z, Li X, Chen H (2005) Link prediction approach to collaborative
  filtering.
\newblock In: Proceedings of the 5th ACM/IEEE-CS Joint Conference on Digital
  Libraries. ACM, pp. 141--142.
\bibAnnoteFile{huang2005link}

\bibitem{ahn2010link}
Ahn YY, Bagrow JP, Lehmann S (2010) Link communities reveal multiscale
  complexity in networks.
\newblock Nature 466: 761--764.
\bibAnnoteFile{ahn2010link}

\bibitem{latora2001efficient}
Latora V, Marchiori M (2001) Efficient behavior of small-world networks.
\newblock Phys Rev Lett 87: 198701.
\bibAnnoteFile{latora2001efficient}

\bibitem{newman2002assortative}
Newman ME (2002) Assortative mixing in networks.
\newblock Phys Rev Lett 89: 208701.
\bibAnnoteFile{newman2002assortative}

\bibitem{dong2011link}
Dong Y, Ke Q, Wang B, Wu B (2011) Link prediction based on local information.
\newblock In: Proceedings of the International Conference on Advances in Social
  Networks Analysis and Mining. IEEE, pp. 382--386.
\bibAnnoteFile{dong2011link}

\bibitem{adamic2005political}
Adamic LA, Glance N (2005) The political blogosphere and the 2004 us election:
  divided they blog.
\newblock In: Proceedings of the 3rd International Workshop on Link discovery.
  ACM, pp. 36--43.
\bibAnnoteFile{adamic2005political}

\bibitem{von2002comparative}
Von~Mering C, Krause R, Snel B, Cornell M, Oliver SG, et~al. (2002) Comparative
  assessment of large-scale data sets of protein--protein interactions.
\newblock Nature 417: 399--403.
\bibAnnoteFile{von2002comparative}

\bibitem{Pajek}
Bataglj V, Mrvar A. Pajek datasets website.
\bibAnnoteFile{Pajek}

\bibitem{EPA}
\url{http://vlado.fmf.uni-lj.si/pub/networks/data/mix/mixed.htm}.
\bibAnnoteFile{EPA}

\bibitem{spring2002measuring}
Spring N, Mahajan R, Wetherall D (2002) Measuring isp topologies with
  rocketfuel.
\newblock ACM SIGCOMM Computer Communication Review 32: 133--145.
\bibAnnoteFile{spring2002measuring}

\bibitem{leskovec2010predicting}
Leskovec J, Huttenlocher D, Kleinberg J (2010) Predicting positive and negative
  links in online social networks.
\newblock In: Proceedings of the 19th International Conference on World Wide
  Web. ACM, pp. 641--650.
\bibAnnoteFile{leskovec2010predicting}

\bibitem{leskovec2010signed}
Leskovec J, Huttenlocher D, Kleinberg J (2010) Signed networks in social media.
\newblock In: Proceedings of the SIGCHI Conference on Human Factors in
  Computing Systems. ACM, pp. 1361--1370.
\bibAnnoteFile{leskovec2010signed}

\bibitem{Lederberg}
\url{http://vlado.fmf.uni-lj.si/pub/networks/data/cite/default.htm}.
\bibAnnoteFile{Lederberg}

\bibitem{hanley1982ROC}
Hanely JA, McNeil BJ (1982) The meaning and use of the area under a receiver
  operating characteristic (roc) curve.
\newblock Radiology 143: 29--36.
\bibAnnoteFile{hanley1982ROC}

\bibitem{herlocker2004evaluating}
Herlocker JL, Konstan JA, Terveen LG, Riedl JT (2004) Evaluating collaborative
  filtering recommender systems.
\newblock ACM Transactions on Information Systems 22: 5--53.
\bibAnnoteFile{herlocker2004evaluating}

\bibitem{shannon2001mathematical}
Shannon CE (2001) A mathematical theory of communication.
\newblock ACM SIGMOBILE Mobile Computing and Communications Review 5: 3--55.
\bibAnnoteFile{shannon2001mathematical}

\bibitem{cover2012elements}
Cover TM, Thomas JA (2012) Elements of Information Theory.
\newblock John Wiley \& Sons.
\bibAnnoteFile{cover2012elements}

\end{thebibliography}

\end{document}